\documentstyle[preprint,aps,epsfig]{revtex}

\newcommand{\lsim}{\raisebox{0.3mm}{\em $\, <$} 
\hspace{-3.3mm} \raisebox{-1.8mm}{\em $\sim \,$}}
\newcommand{\gsim}{\raisebox{0.3mm}{\em $\, >$}
\hspace{-3.3mm} \raisebox{-1.8mm}{\em $\sim \,$}}

\begin{document}
\draft
\title{Constraining Almost Degenerate Three-Flavor Neutrinos} 

\author{Hisakazu Minakata{$^{1,2}$} and Osamu Yasuda{$^1$}}

\address{\qquad \\ {$^1$}Department of Physics,
Tokyo Metropolitan University \\
Minami-Osawa, Hachioji, Tokyo 192-03, Japan\\
\qquad\\{$^2$}Institute for Nuclear Theory, University of Washington \\
Seattle, Washington 98195-1550, USA\\}

\date{September, 1996}

\preprint{
\parbox{5cm}{
TMUP-HEL-9610\\
hep-ph/9609276\\
}}

\maketitle
\begin{abstract}
We discuss constraints on a scenario of almost degenerate
three-flavor neutrinos imposed by the solar and the atmospheric
neutrino anomalies, hot dark matter, and neutrinoless double $\beta$
decays. It is found that in the Majorana version of the
model the region with relatively large $\theta_{13}$ is favored and
a constraint on the CP violating phases is obtained.
\end{abstract}
\vfill

There exist several experimental hints which indicate that most 
probably neutrinos have tiny masses and flavor mixings. The first is 
the solar neutrino deficit observed in four different experiments, 
the chlorine, the Kamiokande II-III, GALLEX and SAGE 
\cite {Davis,Kam,GALLEX,SAGE}. 
It became highly unlikely that the data of various experiments can be 
reconciled with any sensible modifications of the standard solar 
model. The second is the atmospheric neutrino anomaly, the large 
deviation in the observed ratio $\nu_\mu / \nu_e$ from the expectation 
of the Monte-Carlo simulations \cite {Kam2,IMB}. While the anomaly 
was not observed within the statistics of the NUSEX and the Frejus 
experiments \cite {nusex,frejus}, the evidences in the Kamiokande and 
IMB detectors are so impressive that they force us to consider seriously 
about the anomaly. The presence of the anomaly is also supported by 
the newest tracking detector, Soudan 2 \cite {soudan2}.

The possible third hint for neutrino masses comes from the
cosmological model with cold and hot dark matter (CHDM). The neutrinos
is the only known candidate for the hot component. They could be
responsible for the large-scale structure formation in a way
consistent with the COBE observation of anisotropy of cosmic microwave
background \cite {chdm23,ps,bss}. While less direct compared with the
first and the second hints, it provides a good motivation for
examining the possibility of neutrino masses of a few eV range. 

It has been pointed out by various authors that if at least one of the
neutrino states has mass of the dark matter scale and if there is a
hierarchy in two $\Delta m^2$, the difference in squared masses, the
accelerator and the reactor experiments put powerful constraints on
mixing angles \cite {mina,BBGK,FLS}. It is very remarkable that the
mixing pattern of neutrinos is determined to be essentially unique
\cite {mina} if one imposes the additional constraints that come
from the requirement of solving either the solar neutrino problem or
the atmospheric neutrino anomaly, together with that from
neutrinoless double $\beta$ decays \cite{beta}.

The problem with the above framework with only three-flavor neutrinos
(i.e., without sterile neutrinos) is that one cannot account for the
solar neutrino deficit, the atmospheric neutrino anomaly, and the hot
dark matter simultaneously. The only known possibility that can
accommodate these two phenomena as well as supplying neutrino masses
appropriate for hot dark matter within the standard three-flavor
framework is the case of almost degenerate neutrinos (ADN). An
incomplete list of earlier references on ADN is in \cite {bss,ADN}.

In this paper we discuss the constraints that can be imposed on such 
almost degenerate neutrino scenario from the solar and the atmospheric 
neutrino observations as well as the terrestrial neutrino experiments. 
We will point out that, in the case of Majorana neutrinos, the 
neutrinoless double $\beta$ decay experiment \cite{beta} is of key 
importance. In particular, the solar neutrino and the double $\beta$ 
decay experiments constrain the mixing angle $\theta_{13}$ to be large.

Let us start by defining more precisely what we mean by the almost
degenerate neutrinos. Due to the requirement of solving the solar and
the atmospheric neutrino problems the two $\Delta m^2$ should have
values $\lsim 10^{-5}$eV$^2$ and $\sim 10^{-2}$eV$^2$,
respectively. This implies that three neutrino states are degenerate
up to the accuracy of 0.1 eV. Then, the requirement from the hot dark
matter hypothesis implies that they must have masses of the order of a
few to several eV \cite {chdm23,ps,bss}. Then, the degeneracy in
masses is smaller than 0.01eV, hence the name of almost degenerate
neutrinos (ADN).

For definiteness, we assign the smaller $\Delta m^2$ to 
$\Delta m^2_{12} \equiv m_2^2 - m_1^2$ and the larger to 
$\Delta m^2_{13}$. It should be noticed that this can be done without 
loss of generality. Despite the almost degeneracy in neutrino masses 
there is a hierarchy in $\Delta m^2; 
\Delta m^2_{13} \simeq \Delta m^2_{23} \gg \Delta m^2_{12}$. 
It allows us to simplify greatly formulae for the oscillation 
probabilities. With neutrino mixing matrix $U_{\alpha i}$ they read, 
\begin{eqnarray}
P(\nu_\beta \rightarrow \nu_\alpha) &=&
4|U_{\alpha3}|^2|U_{\beta3}|^2 
\sin^2(\displaystyle\frac{\Delta m^2_{13}L}{4E})
\label{eq:osci1}
\\
1-P(\nu_\alpha \rightarrow \nu_\alpha) &=& 
4|U_{\alpha3}|^2(1-|U_{\alpha3}|^2) \sin^2(\displaystyle
\frac{\Delta m^2_{13}L}{4E}),
\label{eq:osci2}
\end{eqnarray}
where the CP violating terms have been dropped in the approximation
with the mass hierarchy, i.e., they are obtained under the so called 
one mass scale dominance approximation.

We use the standard form of Cabibbo-Kobayashi-Maskawa quark mixing matrix 

\begin{eqnarray}
U=\left[
\matrix {c_{12}c_{13} & s_{12}c_{13} &  s_{13}e^{-i\delta}\nonumber\\
-s_{12}c_{23}-c_{12}s_{23}s_{13}e^{i\delta} & 
c_{12}c_{23}-s_{12}s_{23}s_{13}e^{i\delta} & s_{23}c_{13}\nonumber\\
s_{12}s_{23}-c_{12}c_{23}s_{13}e^{i\delta} & 
-c_{12}s_{23}-s_{12}c_{23}s_{13}e^{i\delta} & c_{23}c_{13}\nonumber\\}
\right],
\label{eqn:CKM}
\end{eqnarray}
for the neutrino mixing matrix. From (\ref{eq:osci1}) and (\ref{eq:osci2}) 
one can see that the accelerator and the reactor experiments probe 
the mixing angles $\theta_{13}$ and $\theta_{23}$. 

We first summarize the constraints from the accelerator and the reactor 
experiments. Unlike the case of dark-matter-mass neutrinos with hierarchy 
the constraints form these terrestrial experiments are very mild. 
With $\Delta m^2 \lsim 0.01$eV$^2$ only the relevant channel is $\nu_e$ 
disappearance experiments whose most extensive runs were done at Bugey 
\cite {Bugey} and at Krasnoyarsk \cite {Kras}. 
We note that there is no constraint on $s_{13}^2$ for $\Delta m_{13}^2
< 7\times 10^{-3}$eV$^2$, where we have made the substitutions of the
variables in \cite {Bugey} and \cite {Kras}
$\theta\rightarrow\theta_{13},~\Delta m^2\rightarrow\Delta m^2_{13}$,
which follow from the present approximation with the mass hierarchy. 
Notice that there is an allowed region at large
$s_{13}^2$, whose dominant part will be excluded by the solar neutrino
constraint as we will see below.

Let us now address the constraint from the solar neutrino experiments. 
While extensive analyses have been done within two-flavor mixing 
scheme the full three-flavor analysis of the solar neutrino 
experiments is very rare. To our knowledge it has been carried out 
quite recently for the MSW mechanism \cite {MSW} by Fogli, Lisi, and 
Montanino \cite {FLM}. We do not know corresponding analyses for 
the vacuum oscillation solution, while there was an attempt \cite {BR}.   
For this reason let us focus on the MSW solution in this paper. 

In their extensive analysis in the three-flavor framework with
mass-squared hierarchy they observed amazingly that the MSW solution
still exists at large $s_{13}^2$ \cite {FLM}, contrary to what was
naively thought.  The well-known small-$s_{12}$ and large-$s_{12}$
solutions fuse into a single one at around $s_{13}^2=0.33$ and this
large-$s_{13}$ solution extends up to $s_{13}^2 \simeq 0.6$. The
large-$s_{13}$ solution is interesting because the ``two-flavor''
parameters $\Delta m_{12}^2$ and $s_{12}^2$ differ from that obtained
by the two-flavor analysis. At the largest value of $s_{13},
s_{13}^2=0.6$, $s_{12}^2 \simeq 2\times 10^{-2}$ and $\Delta m_{12}^2
\simeq 4\times 10^{-6}$eV$^2$. In contrast, the best fit values of the
two-flavor analysis are $s_{12}^2 \simeq 2\times 10^{-3}$ and $\Delta
m_{12}^2 \simeq 5.2\times 10^{-6}$eV$^2$. We will see below that the
constraint from neutrinoless double $\beta$ decays does indeed prefer
the large-$s_{13}$ solution.

We discuss the constraint from neutrinoless double $\beta$ decays, which 
applies only to the Majorana neutrinos. We will see that it gives 
rise to the strongest constraint. 
Observation of no neutrinoless double $\beta$ decay implies the 
constraint on $<m_{\nu e}>$, which can be written in our notation 
of the mixing matrix as 
\begin{equation}
<m_{\nu e}> = \left\vert c_{12}^2c_{13}^2 m_1 e^{-i(\beta+\gamma)}
+ s_{12}^2c_{13}^2 m_2 e^{i(\beta-\gamma)}
+ s_{13}^2 m_3 e^{2i(\gamma-\delta)}
\right\vert,
\label{eq:beta1}
\end{equation}
where $\beta$ and $\gamma$ are the extra CP-violating phases 
characteristic to Majorana neutrinos \cite {SV,FY}. 

Let us first discuss the CP-invariant case 
($e^{2i\beta}, e^{i(\beta+3\gamma-2\delta)}=\pm 1$) 
because it is easier to understand. 
In this case the phase factors in (\ref{eq:beta1}) can be
reduced to the CP parities $\eta_j$ of mass eigenstates $j$ with
masses $m_j$ \cite {Wol}. Under the circumstance of almost degeneracy
with which we are working we can approximate the expressions of
$<m_{\nu e}>$ by ignoring the mass differences. Then, it further
simplifies depending upon the pattern of the CP parities of three
neutrinos. Let us take the convention that $\eta_1= +$ and denote them
collectively as $(\eta_1, \eta_2, \eta_3)\equiv
(1,e^{2i\beta},e^{i(\beta+3\gamma-2\delta)}) =(+ + -)$ etc.
Then,
\begin{equation}
\label{eq:beta2}
r \equiv\frac{<m_{\nu e}>}{m}=
\left\{
\begin{array}{ll}
1 \;\;\;& \mbox{for} \;\;\;\;(+++) \\
\left\vert 1-2s_{13}^2 \right\vert\;\;\;& \mbox{for} \;\;\;\;(++-) \\
\left\vert 1-2s_{12}^2c_{13}^2 \right\vert\;\;\;& \mbox{for} \;\;\;\;(+-+) \\
\left\vert 1-2c_{12}^2c_{13}^2\right\vert \;\;\;& \mbox{for} \;\;\;\;(+--)
\end{array}
\right.
\end{equation}
Let us refer to the ratio $<m_{\nu e}>/m$ as $r$ hereafter. 

We take the mixed dark matter model with $\Omega_{total}=1$ to 
estimate the masses of neutrinos. The CHDM model with 
$\Omega_{total}=1$ might have problems with age of the universe. 
The measurement by the Hubble Space Telescope \cite{hst} gave a 
value of $h=0.8\pm0.17$, where the Hubble constant $H_0$ is given by 
$h$ as $H_0$=100 $h$ km/s $\cdot$ Mpc. The value of $H_0$ suggests 
that the total contribution $\Omega_{total}$ by matter to the 
density parameter should be smaller than 1 in order to have the age 
of the universe greater than 10 Gyr. 
Our attitude to this problem is that we must take at least 2 $\sigma$ 
uncertainty in the observed value of the Hubble constant seriously 
because the systematic errors in various methods of measuring the 
Hubble constant do not appear to be well understood. (For a recent 
status of measurement of the Hubble constant, see, e.g., \cite {Schramm}). 

We assume that 20-30\% of the universe is shared by the hot dark 
matter. We note that the neutrino contribution to the $\Omega$ 
parameter is $\Omega_{\nu}$ = $ (\sum m_{i}/ 91.5\mbox{eV})h^{-2}$ 
\cite {Kolb}. The CHDM model with three kinds of neutrinos has been 
analyzed by Pogosyan and Starobinsky \cite{ps} and they concluded that 
the allowed region is given by $0.55\lsim h\lsim0.7$,
$0.25\lsim\Omega_\nu\lsim0.3$.  If we take these values, we obtain
$2.3$ eV$\lsim m_j \simeq m \lsim 4.5 $eV as masses of almost degenerate
neutrinos. We will use $m$=2.3 eV, and 4.5 eV for neutrino mass as
reference values in the following analysis.

We impose the experimental bound on $<m_{\nu e}>$ obtained by negative
results of neutrinoless double $\beta$ decays. The most stringent one
to date is $<m_{\nu e}> \lsim 0.68$eV derived in the $^{76}$Ge
experiment by Klapdor-Kleingrothaus et al. as quoted in
\cite{beta}. It implies the bound on the $r$ parameter 
$r \leq$ 0.29, and 0.15 for neutrino masses $m$=2.3 eV, and 4.5 eV,
respectively.

The immediate consequence of the constraint from neutrinoless double
$\beta$ decays is that the first pattern of the CP parity, $(+ + +)$,
is excluded. Other patterns are not immediately excluded but their
parameters are subject to the constraint. In Figs. 1 and 2 we have
plotted the allowed regions of the neutrinoless double $\beta$ decay
constraint for each respective pattern of the CP parities for
$r\le$ 0.15 and $r\le$ 0.29, respectively.  The upper rectangular region
is for the CP parity $(+ + -)$, the lower-left band for $(+ - -)$, and
the lower-right band for $(+ - +)$.  Also plotted in Figs. 1 and 2 as
darker shaded parts are the allowed regions with 90\% CL for the
three-flavor MSW solution to the solar neutrino problem obtained by
Fogli et al. \cite {FLM}. It is actually the superposition of the
allowed regions with the mass squared difference $\Delta m_{12}^2$
from $10^{-6}$eV$^2$ to $1.0\times10^{-4}$eV$^2$.

From this figure one can draw several conclusions: The CP parity
pattern of $(+ - +)$ is excluded since the two allowed regions do not
overlap. The patterns $(+ + -)$ and $(+ - -)$ are allowed and they
prefer the large-$s_{13}$ solution of solar neutrino problem. The
``two-flavor'' large angle solution is also marginally allowed.  The
small angle MSW solutions, which are drawn almost on the axis of
$s_{12}^2=0$ in Figs. 1 and 2, are not compatible with the double
$\beta$ decay constraint for $m$=2.3 eV and 4.5 eV.  In closer detail,
with neutrino mass of 4.5eV ($r\le$ 0.15) the solution exists for
3.2$\times10^{-6}$eV$^2\lsim \Delta m_{12}^2\lsim
6.8\times10^{-5}$eV$^2$ only if $s_{13}^2\gsim0.3$. With $m$=2.3eV
($r\le$ 0.29) it exists for 3.2$\times10^{-6}$eV$^2\lsim \Delta
m_{12}^2\lsim 1.0\times10^{-4}$eV$^2$ only if $s_{13}^2\gsim0.02$.

In general CP-noninvariant cases, we have to keep the two CP violating
phases $\beta$ and $\gamma$
in (\ref{eq:beta1}). Namely, we have
\begin{eqnarray}
<m_{\nu e}> &=& m \left\vert c_{13}^2(e^{-i\beta}c_{12}^2
+ e^{i\beta} s_{12}^2)
+ e^{i(3\gamma - 2\delta)} s_{13}^2 \right\vert\nonumber\\
&\ge&m\left\vert 
c_{13}^2(1-\sin^2\beta\sin^22\theta_{12})^{1/2}
-s_{13}^2\right\vert,
\label{eq:beta3}
\end{eqnarray}
where we have ignored the mass differences and the equality in the
second line holds when
\begin{eqnarray}
\arg (e^{-i\beta}c_{12}^2+e^{i\beta}s_{12}^2)=
3\gamma - 2\delta+(2n+1)\pi,
\label{eq:phases}
\end{eqnarray}
where $n$ is an integer.  Note that the constraint from neutrinoless
double $\beta$ decays becomes even more stringent if the CP violating
phases $\beta, \gamma, \delta$ do not satisfy the relation
(\ref{eq:phases}).  Then, our task is to look for the region which
satisfies $r \leq$ 0.29, 0.15, respectively, with $\beta$,
$\gamma$, and $\delta$ unconstrained.  The resulting bounds coincide
with those obtained in the CP-conserving cases and are
presented in Figs. 3 and 4.  For $r\le$ 0.15 which is obtained from $h$=0.7 and
$\Omega_\nu$=0.3, the solution exists only if $s_{13}^2\gsim 0.3$, and
for $r\le$ 0.29 only if $s_{13}^2 \gsim 0.02$.
It should be emphasized that, irrespective of whether CP is violated
or not, the small-$s_{12}$ MSW solution, which is favored by theorists
most, is disfavored in the CHDM model with almost degenerate neutrino
masses.

On the other hand, we can get a condition for the CP violating phases
$\beta$, $\gamma$, and $\delta$ by imposing both constraints from
neutrinoless double $\beta$ decays and from the solar neutrino
deficits with $s_{12}^2$ and $s_{13}^2$ constrained by the
three-flavor analysis of \cite {FLM}.  The results are shown in Figs. 
5 and 6, where the allowed regions are located in the neighborhood of
the line $\beta+3\gamma-2\delta=\pm\pi$.  In these plots the
CP-conserving cases with the patterns $(+ + +)$, $(+ + -)$, $(+ - +)$
and $(+ - -)$ correspond to the points $(\beta,3\gamma-2\delta)$ =
(0,0), $\pm(\pi,\pi)$, $\pm(\pi,-\pi)$; (0,$\pm\pi$), ($\pm\pi$,0);
$\pm(\pi/2,-\pi/2)$; $\pm(\pi/2,\pi/2)$, respectively.  From this we
can verify in Figs. 5 and 6 that the CP-conserving cases with the CP
parities $(+ + +)$ and $(+ - +)$ are indeed excluded both for $r\le$
0.15 and $r\le$ 0.29.

Finally, let us briefly discuss the constraints from the atmospheric
neutrino anomaly. There have been several three-flavor analyses of the
atmospheric neutrino anomaly \cite {atm3,Yasuda,FLMS}. Among them, the
most recent and the most detailed are the ones done by one of the
authors \cite {Yasuda}, and by Fogli, Lisi, Montanino, and Scioscia
\cite {FLMS}. The analyses by these two groups give rise to slightly
different 2$\sigma$ allowed regions on the $s_{13}^2 - s_{23}^2$
parameter plane. The difference stems from their treatments of the
data of the NUSEX \cite{nusex} and Frejus \cite{frejus} experiments
which are included in \cite{FLMS} and are not in \cite{Yasuda}.  As
far as the constraint for $\theta_{23}$ is concerned, it is concluded
in either analysis \cite{Yasuda,FLMS} that the allowed region with 90
\% CL for $\Delta m_{13}^2\sim 5\times10^{-3}$ eV$^2$ has to satisfy
$s_{23}^2\gsim 1/4$.  However, there is a
difference between the two analyses on the allowed region for
$\theta_{13}$.  If one includes the data of all
the experiments of atmospheric neutrinos \cite{FLMS}, then the
solution with small-$s_{13}$ is allowed.  On the other hand, if one
considers only the multi-GeV Kamiokande data \cite{Yasuda}, the
solution with $s_{13}^2\lsim0.1$ is excluded at 90 \% confidence level.
As we have seen above, the allowed region for $r\le$ 0.15
exists for rather large values of $s_{13}$, so the difference of the
two analyses \cite{Yasuda,FLMS} turns out to be irrelevant for
$m\gsim 3$eV.

To summarize, we have discussed the almost degenerate three-flavor
neutrino scenario as a simultaneous solution to the solar, the
atmospheric and the dark matter problems. We have shown, using the
constraints from neutrinoless double beta decays as well as these
observational data of the solar and the atmospheric neutrinos, that
large value of $s_{13}^2$ is favored, leaving a little room for
solutions with small $s_{13}^2$ and large $s_{12}^2$. The neutrinoless
double $\beta$ decay constraint imposed in ADN makes the small angle
MSW solution untenable in this scenario. If three neutrinos turn out
to be degenerate in masses and if precise values of $s_{12}^2$ and
$s_{13}^2$ are both determined experimentally, then we get information
on the relation among the CP violating phases $\beta$, $\gamma$ and
$\delta$.

The authors would like to thank N. Okada and S. Sasaki for discussions
on astrophysical aspects of dark matter scenarios. One of us (H.M.) is 
grateful to George Fuller for illuminating conversations on CHDM cosmology, 
and to Institute for Nuclear Theory, University of Washington for 
partial support during the completion of this work.
We have been supported in part by Grant-in-Aid for Scientific Research 
of the Ministry of Education, Science and Culture under \#0560355, 
and H.M. is also supported by Grant-in-Aid for Scientific Research 
on Priority Areas under \#08237214.

\newpage

\vspace{1.5cm}
\centerline{\large FIGURE CAPTIONS}
\vspace{0.5cm}

\begin{description}

\item[Fig.1,2.]
\begin{minipage}[t]{13cm}
\baselineskip=20pt

The Lighter shaded areas bounded by the thicker lines are the allowed
regions of the neutrinoless double $\beta$ decay constraints for each
respective pattern of the CP parities in the CP-conserving case with
$r<$ 0.15 ($m$=4.5eV) and $r<$ 0.29 ($m$=2.3eV), respectively.  The
darker shaded areas bounded by the thinner lines are the allowed
regions with 90\% CL for the three-flavor MSW solution of the solar
neutrino problem obtained by Fogli et al. \cite {FLM}.
\end{minipage}
\vspace{0.5cm}

\item[Fig.3,4.]
\begin{minipage}[t]{13cm}
\baselineskip=20pt
The Lighter shaded areas are the allowed regions of the neutrinoless
double $\beta$ decay constraints for the general CP
violating case with $r<$ 0.15 ($m$=4.5eV) and $r<$ 0.29 ($m$=2.3eV),
respectively, where the CP violating phases $\beta$, $\gamma$ and
$\delta$ are unconstrained.  The darker shaded areas are the same as in
Figs. 1 and 2.
\end{minipage}
\vspace{0.5cm}

\item[Fig.5,6.]
\begin{minipage}[t]{13cm}
\baselineskip=20pt
The shaded areas are the allowed regions obtained from the
neutrinoless double $\beta$ decay experiments and the solar neutrino
analysis with 90\% CL by Fogli et al. \cite {FLM} for $r<$ 0.15
($m$=4.5eV) and $r<$ 0.29 ($m$=2.3eV), respectively, where $s_{12}^2$
and $s_{13}^2$ take all possible values within the constraint
of \cite {FLM}.
\end{minipage}

\end{description}
\newpage
\pagestyle{empty}
\epsfig{file=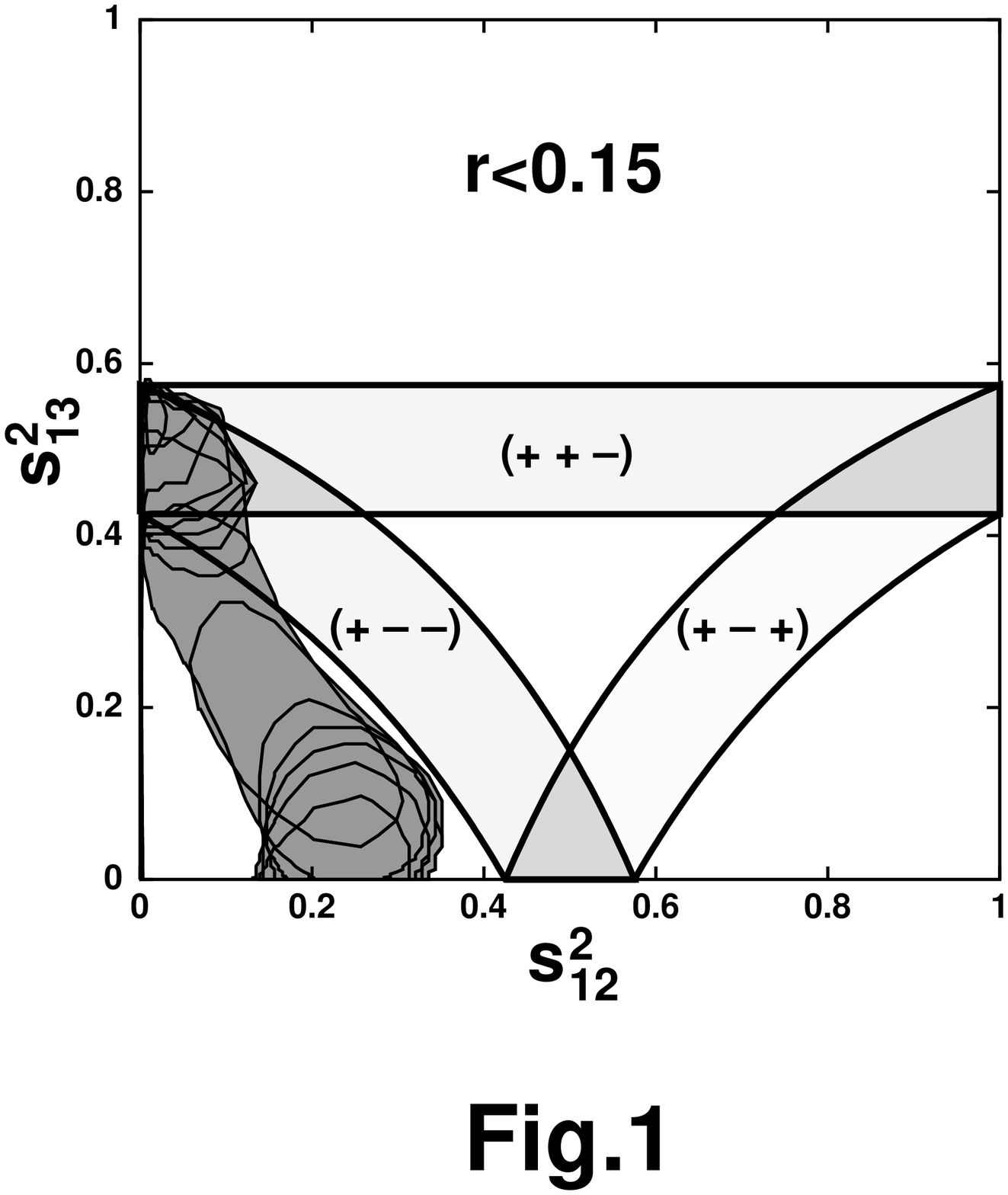,width=15cm}
\newpage
\epsfig{file=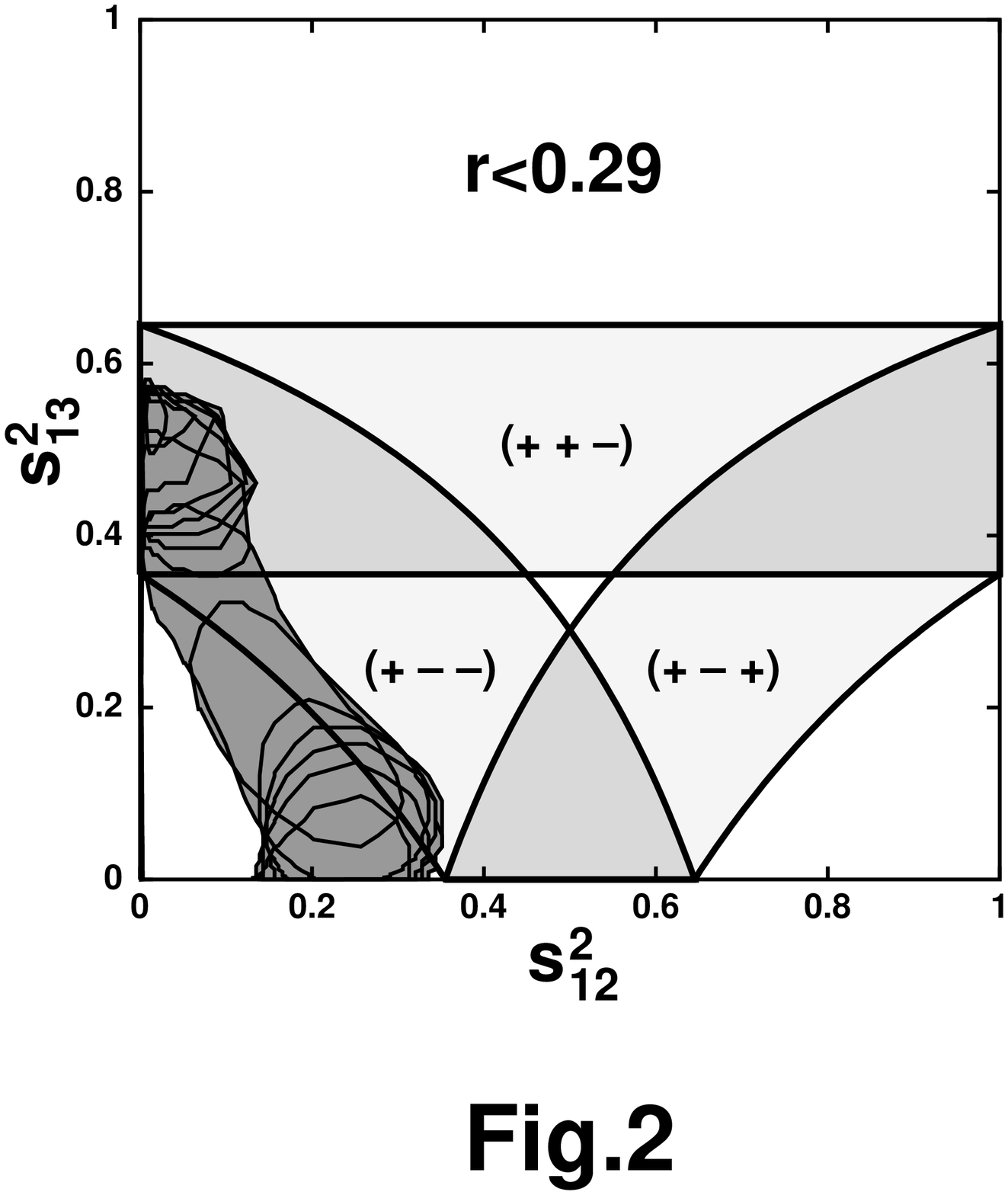,width=15cm}
\newpage
\epsfig{file=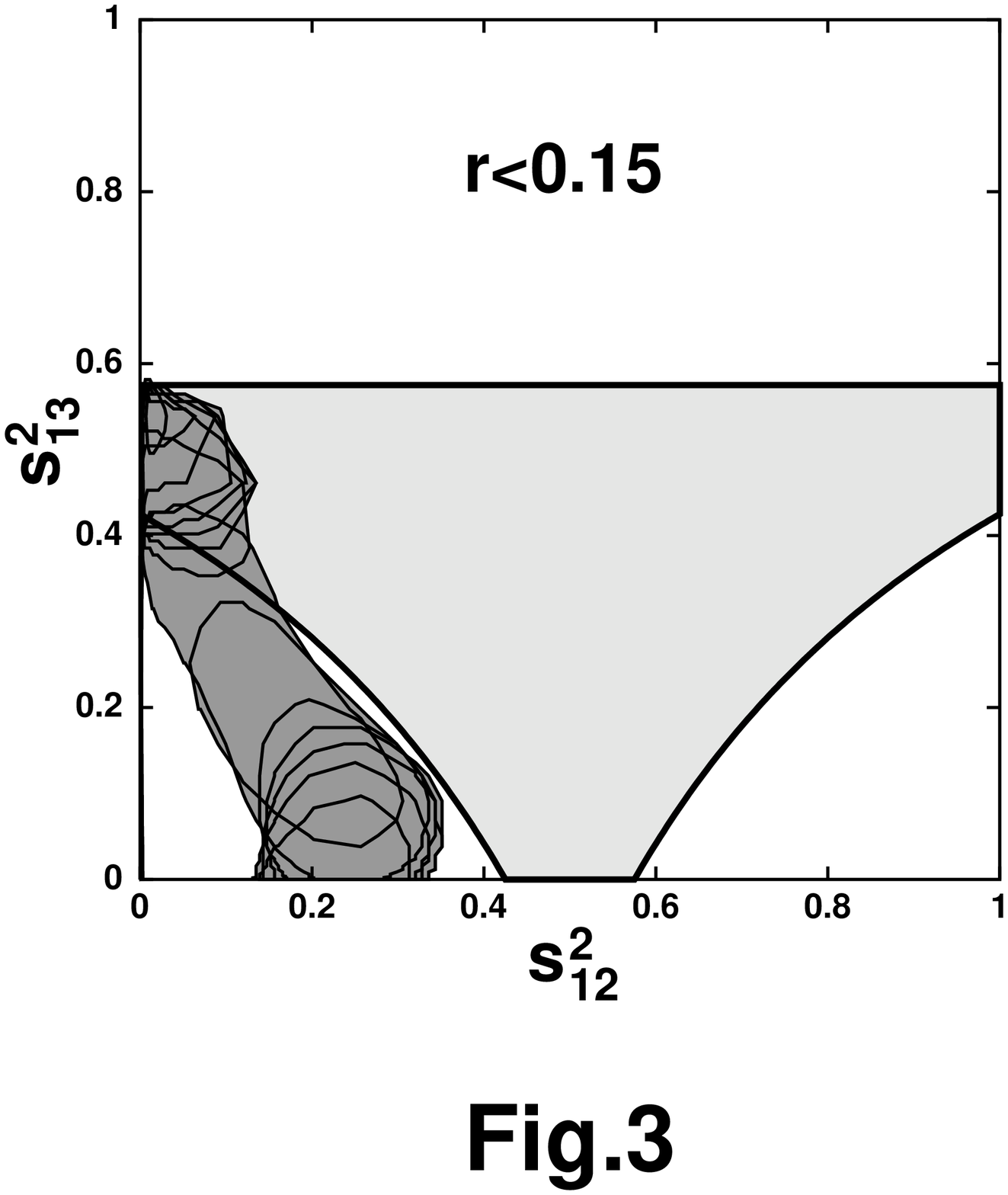,width=15cm}
\newpage
\epsfig{file=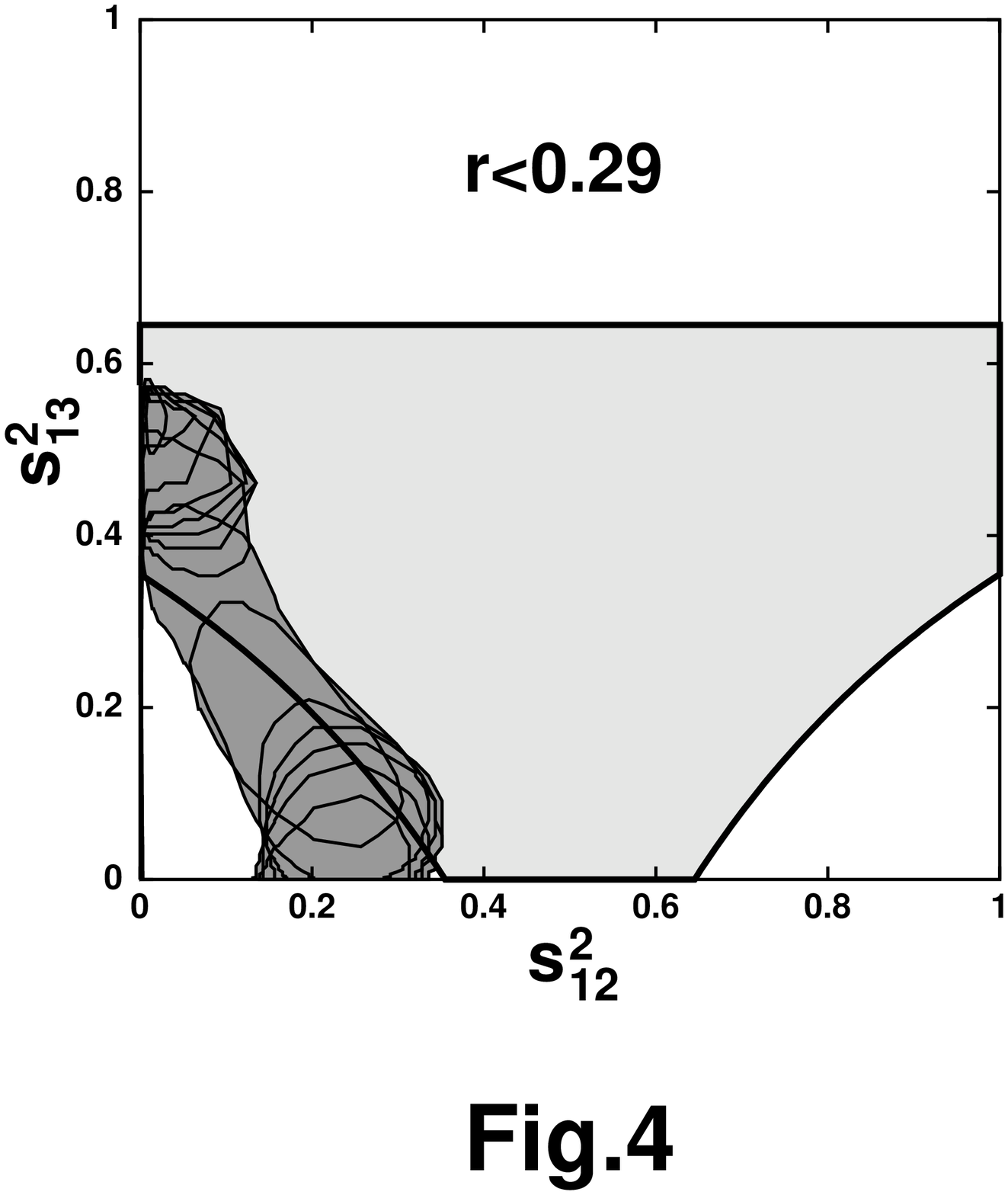,width=15cm}
\newpage
\epsfig{file=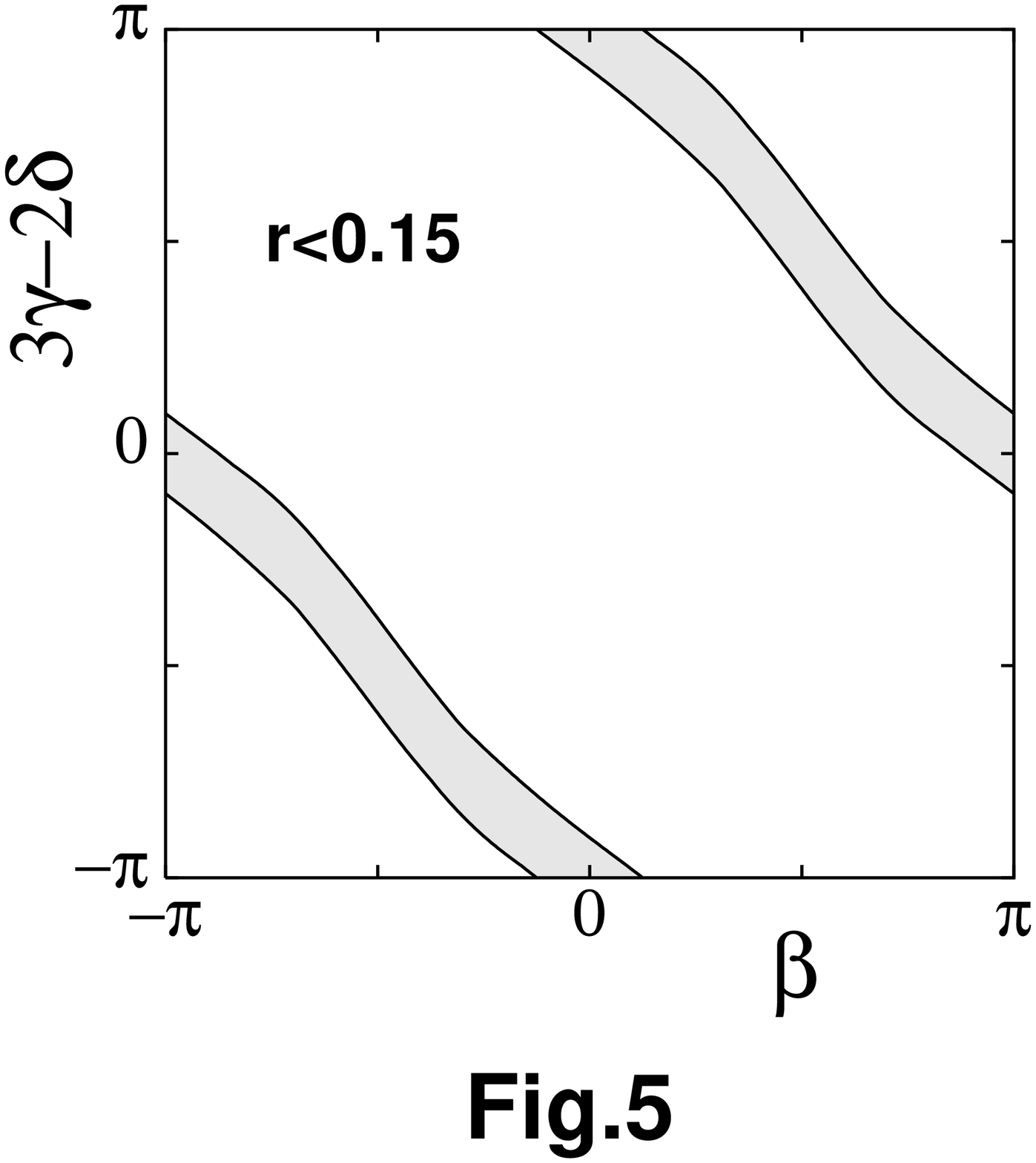,width=15cm}
\newpage
\epsfig{file=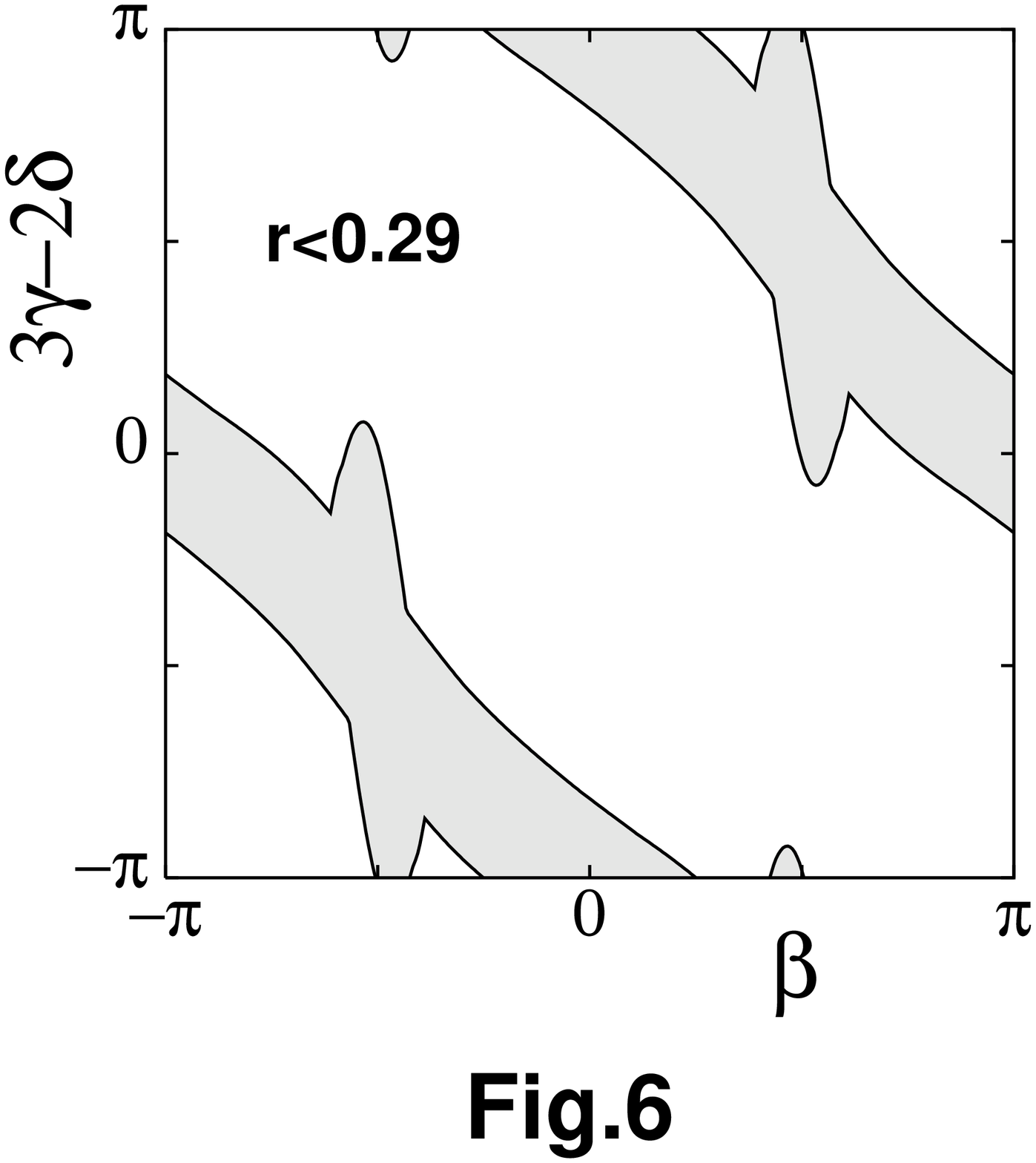,width=15cm}


\begin{references}

\bibitem{Davis}
B. T. Cleveland et al., Nucl. Phys. B (Proc. Suppl.) {\bf 38} (1995) 47.

\bibitem{Kam}
K. S. Hirata et al., Phys. Rev. {\bf D44} (1991) 2241;
Y. Suzuki, Nucl. Phys. B (Proc. Suppl.) {\bf 38} (1995) 54.

\bibitem{GALLEX}
P. Anselmann et al., Phys. Lett. {\bf B285} (1992) 376;
ibid {\bf B314} (1993) 445; ibid {\bf 327}, (1994) 377;
ibid {\bf B342}, (1995) 440. 

\bibitem{SAGE}
A. I. Abazov et al., Phys. Rev. Lett. {\bf 67} (1991) 3332;
J. N. Abdurashitov et al., Nucl. Phys. B (Proc. Suppl.) {\bf 38} (1995) 60.

\bibitem{Kam2}
K. S. Hirata et al., Phys. Lett. {\bf B205} (1988) 416;
ibid {\bf B280} (1992) 146; 
Y. Fukuda et al., Phys. Lett. {\bf B335} (1994) 237;

\bibitem{IMB}
D. Casper et al., Phys. Rev. Lett. {\bf 66} (1991) 2561;
R. Becker-Szendy et al., Phys. Rev. {\bf D46} (1992) 3720.

\bibitem {nusex}
M. Aglietta et al., Europhys. Lett. {\bf 8(7)} (1989) 611;

\bibitem {frejus}
K. Daum et al., Z. Phys.  {\bf C66} (1995) 417.

\bibitem {soudan2}
M. C. Goodman, Nucl. Phys. B (Proc. Suppl.) {\bf 38} (1995) 337.

\bibitem {chdm23}
J. A. Holtzman, Astrophys. J. Suppl. {\bf 71} (1989) 1;
J. A. Holtzman and J. R. Primack,
Astrophys. J. {\bf 405} (1993) 428;
J. R. Primack, J. Holtzman, A. Klypin, and D. O. Caldwell,
Phys. Rev. Lett. {\bf 74} (1995) 2160.

\bibitem {ps}
D. Pogosyan and A. Starobinsky, astro-ph/9502019.

\bibitem {bss}
K. S. Babu, R. K. Schaefer, and Q. Shafi,  Phys. Rev. {\bf D53} (1996) 
606.

\bibitem {mina}
H. Minakata, Phys. Rev. {\bf D52} (1995) 6630; Phys. Lett. {\bf B356} 
(1995) 61.

\bibitem {BBGK}
S. M. Bilenky, A. Bottino, C. Giunti, and C. W. Kim, Phys. Lett. 
{\bf B356} (1995) 273.

\bibitem {FLS}
G. L. Fogli, E. Lisi, and G. Scioscia, Phys. Rev. {\bf D52} (1995) 5334.

\bibitem {beta}
M. K. Moe, Nucl. Phys. B (Proc. Suppl.) {\bf 38} (1995) 36.

\bibitem {Bugey}
B. Achkar et al., Nucl. Phys. {\bf B434} (1995) 503. 

\bibitem {Kras}
G. S. Vidyakin et al., JETP Lett. {\bf 59} (1994) 390.

\bibitem {ADN}
D. Caldwell and R. N. Mohapatra, Phys. Rev. {\bf D48} (1993) 3259; 
{\bf D50} (1994) 3477;
S. T. Petcov and A. Yu. Smirnov, Phys. Lett. {\bf B322} (1994) 109;
A. S. Joshipura, Z. Phys. {\bf C64} (1994) 31;
A. Ioannissyan and J. W. F. Valle, Phys. Lett. {\bf B332} (1994) 93;
B. Kayser, Talk at IInd Rencontres du Vietnam, ``Physics at the Frontiers 
of the Standard Model'', Ho Chi Minh, October 21-28, 1995.  

\bibitem{MSW}
S. P. Mikheyev and A. Smirnov, Nuovo Cim. {\bf 9C} (1986) 17;
L. Wolfenstein, Phys. Rev. {\bf D17} (1978) 2369.

\bibitem {FLM}
G. L. Fogli, E. Lisi, and D. Montanino,
INSSNS-AST 96/21, hep-ph/9605273.

\bibitem{BR}
Z. G. Berezhiani and A. Rossi, Phys. Lett. {\bf B367} (1996) 219.

\bibitem {SV}
J. Schechter and J. W. F. Valle, Phys. Rev. {\bf D22} (1980) 2227;
S. M. Bilenky, J. Hosek, and S. T. Petcov, Phys. Lett. {\bf B94} 
(1980) 495;
M. Doi et al., Phys. Lett. {\bf B102} (1981) 323.

\bibitem {FY}
M. Fukugita and T. Yanagida, in {\it Physics and Astrophysics of Neutrinos} 
(Springer-Verlag, Tokyo, 1994)

\bibitem {Wol}
L. Wolfenstein, Phys. Lett. {\bf B107} (1981) 77.

\bibitem {hst}
W.L. Freedman et al.,  Nature {\bf 371} (1994) 385.

\bibitem {Schramm}
D. N. Schramm, Talk presented at Inauguration Conference of Asia Pacific 
Center for Theoretical Physics, June 4-10, 1996, Seoul, Korea.

\bibitem{Kolb}
E. W. Kolb and M. S. Turner, {\it The Early Universe} (Addison-Wesley
Publishing Co., California, 1990).

\bibitem{atm3}
J.~G.~Learned, S.~Pakvasa, and T.~J.~Weiler, 
Phys.\ Lett.\ B {\bf 207}, 79 (1988);
V.~Barger and K.~Whisnant, 
Phys.\ Lett.\ B {\bf 209}, 365 (1988);
K. Hidaka, M. Honda, and S. Midorikawa, 
Phys.\ Rev.\ Lett.\ {\bf 61}, 1537 (1988);
S.~Midorikawa, M.~Honda, and K.~Kasahara,
Phys.\ Rev.\ D {\bf 44}, R3379 (1991);
J.~G.~Learned, S.~Pakvasa, and T.~J.~Weiler,
Phys. Lett. B {\bf 298}, 149 (1993);
A.~Acker, A.~B.~Balantekin, and F.~Loreti, 
Phys.\ Rev.\ D {\bf 49}, 328 (1994);
J.~Pantaleone, 
Phys.\ Rev.\ D {\bf 49}, R2152 (1994);
G.~L.~Fogli, E.~Lisi, and G.~Scioscia,
Phys.\ Rev.\ D {\bf 52}, 5334 (1995);
S.~M.~Bilenky, C.~Giunti, and C.~W.~Kim,
Astropart.\ Phys.\ {\bf 4}, 241 (1996);
J.~J.~Gomez-Cadenas and M.~C.~Gonzalez-Garcia,
CERN-TH-95-80, hep-ph/9504246;
M.~Narayan, M.~V.~N.~Murthy, 
G.~Rajasekaran, and S.~Uma Sankar,
Phys.\ Rev.\ D {\bf 53}, 2809 (1996);
S. Goswami, K. Kar and A. Raychaudhuri,
CUPP-95-3, hep-ph/9505395.

\bibitem {Yasuda}
O. Yasuda, TMUP-HEL-9603, hep-ph/9602342.

\bibitem {FLMS}
G. L. Fogli, E. Lisi, D. Montanino, and G. Scioscia, 
INSSNS-AST 96/41, hep-ph/9607251

\end{references}
\end{document}